\documentclass[11pt]{article}

\usepackage{lmodern}
\usepackage{tgtermes}
\usepackage[T1]{fontenc}

\usepackage{amsmath,amsfonts,amssymb, systeme}

\usepackage[noblocks]{authblk}

\usepackage[verbose,letterpaper,tmargin=2.54cm,bmargin=2.54cm,lmargin=2.54cm,rmargin=2.54cm]{geometry}

\usepackage[numbers,square,comma,sort&compress]{natbib}

\usepackage{placeins}

\usepackage{xspace}

\usepackage[bookmarks,bookmarksnumbered,colorlinks=true,linkcolor=black,urlcolor=black,citecolor=blue]{hyperref}

\usepackage{setspace}

\usepackage{graphicx}
\usepackage{float}
\usepackage{subcaption}

\usepackage{tabularx}
\usepackage{blkarray}
\usepackage{lscape}
\usepackage[table]{xcolor}
\usepackage{colortbl}
\usepackage{booktabs}
\usepackage{makecell}
\usepackage{array}
\usepackage{multirow}
\usepackage{multicol}
\usepackage{boldline}
\usepackage{threeparttable}
\usepackage{arydshln}
\newcolumntype{L}[1]{>{\raggedright\let\newline\\\arraybackslash\hspace{0pt}}m{#1}}
\newcolumntype{C}[1]{>{\centering\let\newline\\\arraybackslash\hspace{0pt}}m{#1}}
\newcolumntype{R}[1]{>{\raggedleft\let\newline\\\arraybackslash\hspace{0pt}}m{#1}}
\usepackage[graphicx]{realboxes}

\usepackage{tikz}
\usepackage{siunitx}
\usetikzlibrary{patterns}
\usetikzlibrary{decorations.pathreplacing}
\tikzset{>=latex} 
\usepackage{siunitx}
\usepackage{verbatim}
\usetikzlibrary{arrows,calc}
\usepackage{relsize}

\usepackage{lineno}
\usepackage{color}
\definecolor{lightgrey}{rgb}{0.5,0.5,0.5}

\definecolor{red}{rgb}{1,0,0}
\definecolor{green}{rgb}{0,0.6,0}
\definecolor{blue}{rgb}{0,0,0.9}

\usepackage{caption}
\usepackage{natbib}

\title{
Reading Yule in light of the history and present\\ of macroevolution 
}

\author[1,2,3,$\dag$, *]{\small Matt Pennell}
\author[4,$\dag$]{\small Ailene MacPherson}

\affil[1]{\emph{Department of Quantitative and Computational Biology, University of Southern California, USA}}
\affil[2]{\emph{Department of Biological Sciences, University of Southern California, USA}}
\affil[3]{\emph{Department of Computational Biology, Cornell University, USA}}
\affil[4]{\emph{Department of Mathematics, Simon Fraser University, Canada}}
\affil[$\dag$]{Both authors contributed equally}
\affil[*]{Corresponding author: mpennell@usc.edu}

\date{}

\begin{document}

\maketitle



\doublespacing

\vspace{0.75cm}

\begin{abstract}
\noindent Yule's 1925 paper introducing the branching model that bears his name was a landmark contribution to the biodiversity sciences. In his paper, Yule developed stochastic models to explain the observed distribution of species across genera and to test hypotheses about the relationship between clade age, diversity, and geographic range. Here we discuss the intellectual context in which Yule produced this work, highlight Yule's key mathematical and conceptual contributions using both his and more modern derivations, and critically examine some of the assumptions of his work through a modern lens. We then document the strange trajectory of his work through the history of macroevolutionary thought and discuss how the fundamental challenges he grappled with --- such as defining higher taxa, linking microevolutionary population dynamics to macroevolutionary rates, and accounting for inconsistent taxonomic practices --- remain with us a century later. 
\end{abstract}

\newpage


\section*{Introduction}

In 1925, George Udny Yule published a groundbreaking paper titled "A mathematical theory of evolution, based on the conclusions of Dr. J. C. Willis, F.R.S." \citep{Yule1924} in the Philosophical Transactions of the Royal Society. This work introduced the model that later became known as the "Yule process", a stochastic model describing the growth and diversification of genera and species. The impact of this paper extends far beyond the fields we would now recognize as macroevolution, biogeography, and macroecology, influencing the broader use of mathematical models in the life sciences and social sciences. In this perspective, we provide the intellectual context in which Yule developed his approach, explain his model in terms of more contemporary mathematical techniques and notation, discuss the influence his work had (and did not have) on the development of macroevolutionary thought, in particular. (We will touch briefly on other research areas where Yule's paper influenced work, but only where they intersect with macroevolution.) We also highlight how some of the ideas Yule was wrestling with --- the hierarchical nature of biodiversity, the role of taxonomy in shaping patterns of biodiversity, and the effect of demography on speciation --- are directly related to outstanding issues today.

In the early 20th century, evolutionary theory was undergoing a period of intense debate and reevaluation. One of the central controversies during this time was the dispute between proponents of macromutational theories and those who advocated for gradualism. Macromutational theorists, such as de Vries \citep{deVries1901,deVriesMacDougal1905}, argued that new species could arise suddenly through mutations of large effect. Their intellectual opponents argued that the continuous variation observed in natural populations was more consistent with a gradual process of evolutionary change \citep{Provine1971}. It was against this backdrop that J.C. Willis proposed his "Age and Area" hypothesis in his influential book of the same name \citep{Willis1922}. In the book, Willis worked to understand the implications of macromutationism, which he supported, for the distribution of biodiversity across space (de Vries also contributed to the volume). Willis posited that the geographic range of a species is determined by its age, with older genera having wider distributions and younger ones having more restricted ranges. This idea was supported both by first principles reasoning from a macromutationist perspective and on his own observations of plant species distributions, particularly on islands, as well as those of his contemporaries (e.g., \citep{Reid1899, Guppy1903}).

\section*{Yule's models and empirical tests}

Yule was not satisfied by Willis’ reasoning. Rather he proposed that the relationship between the age of a species and its geographic distribution was likely a more complex result of the species' age and its demography. At the time that Willis' book was published, Yule was already an accomplished and influential applied mathematician. (For discussions of his contributions see \citep{Tabery2004,Yates1952}; we note that Yule's work in macroevolution warrant only one line of discussion in ref. \citep{Yates1952}.)
Yule, who had previously co-authored a paper with Willis on the distribution of biodiversity \citep{WillisYule1922}, felt that the Age and Area hypothesis did not make sense in light of Darwinian evolution:
\begin{quote}
\emph{On the Darwinian view that species are continually dying out --- that a species rises,
flourishes and dies, superseded by the more advantageous form --- a species occupying a very small area may be young, but it is equally likely or more likely to be old (a dying species). On Darwin’s own view that the whole body of individuals in a species becomes altered together, the young, species must be found occupying a large area at once, and the species occupying a small area could only be a “dying” species. On the Darwinian view therefore either there need be no relation between Age and Area, or there would be a negative relation, species occupying small areas being on the whole the oldest. Similarly, on the Darwinian view a genus of a few, or of only one, species may be either young or old --- a dying genus --- and there need be no necessary relation between Age and Size. That species occupying very small areas, and the species of monotypic genera are mainly “relic” forms, is, I gather, the predominant Darwinian view. Dr. Willis’s conclusions are inconsistent with that view. [p.22]}
\end{quote}
More importantly, Yule recognized that a more rigorous, quantitative approach was required to generate meaningful hypotheses about the relationship between clade size, age, and geographic distribution.

While there are many derivations of the "Yule model" available (e.g., \cite{Durrett1999,SteelMcKenzie2001}), these are often challenging to square with the actual models presented in his landmark paper. This is because how he formulated the problem, his notation, and his mathematical techniques for solving the problem are all quite distinct from more modern approaches. 
Yule was interested in the relationship between the age of a clade  (which he referred to as a genus) and the number of species in that clade. 
He viewed this as a demographic question involving the dynamics of individuals within a species, species within genera, and birth of new genera through time.  
It was through this demographic lens that he referred to the "age" of a genus and the "birth" and "death" of species (\citep[][p. 22]{Yule1924}).
Intriguingly, Yule's demographic approach does not naturally require any explicit tree structure \citep{lambert2024} (see for example, \citep{simon1955class}).

To bridge the gap between this original conceptualization versus modern uses and to better illustrate how Yule originally worked out his solution, we re-derive his results through a modern lens. We take two complementary approaches. First, we try to faithfully reproduce his modeling results using notation and terminology that will be more familiar to contemporary researchers. Second, we show how his derivations are different from later iterations on the same basic template that relied on innovations in stochastic process theory (following Bailey's 1964 formulation \citep{Bailey1964}) and coalescent thinking (following Nee and colleague's influential papers \citep{NeeEtAl1992, NeeEtAl1994}); we put all three into a common notation so that the essential differences will be more apparent.

We start, as Yule did, with the simplest of scenarios. Yule considered a clade originating from a single "primordial" (\citep[][p. 21]{Yule1924}) lineage at time $t=0$ which diversifies forward in time between the origin and the present day, resulting in a clade of age (or, alternatively, `tree height') $T$. Throughout we will use $t$ to denote forward-time (from the origin $t=0$ to the present day $t=T$) as used by Yule and $\tau$ to denote backward-time (from the present $\tau=0$ to the origin $\tau=T$) as used in coalescent-style approaches; that is, whichever way you count time, the total time covered by the tree is the same. Yule assumed that all lineages are exchangeable \citep{StadlerBonhoeffer2013} (also referred to as a "single-type" model \citep{MacPhersonEtAl2021a}), such that the processes of diversification are identical for all lineages, and that the population sizes of the constituent species do not influence the processes of diversification at all (we will return to both of these assumptions below). Given these assumptions, Yule (\citep[][p. 36]{Yule1924}) considered the expected clade size $\bar{n}(T)$ and the probability that a clade has $n$ lineages $P_n(T)$ at the present day ($t=T$). While the "Yule model" is understood today as a diversification process without extinction, Yule actually had a particular extinction scenario in mind, which he referred to as "cataclysm" (\citep[][p. 80]{Yule1924}). In other words, bad things happen to good lineages --- land masses sink into the ocean, volcanoes erupt, and rivers dry up. These are geologically instantaneous, they occur at random intervals, and are, on average, non-selective. Of course, if a river dries up, all of the species that make their home there will be at risk of extinction, but next time it happens, it will likely be a different river. Yule claimed that if these random events were the predominant mode of extinction, he could safely ignore them in the calculation of clade size. As later work has shown \citep{Kendall1948,Bailey1964,Raup1985,NeeEtAl1994}, this is not actually true and the quantities Yule derived are correct if the only process that changes clade size is speciation.

\subsection*{Yule's probabilistic approach}

To derive the mean and full distribution of clade sizes arising from a given primordial lineage, Yule (\citep[][p. 33]{Yule1924}) divided the total time $T$ into $m$ small intervals of length $\Delta t=\frac{T}{m}$ such that the probability that a lineage speciates in a given unit of time is $p=\lambda\Delta t=\frac{\lambda T}{m}$ . The corresponding probability that a lineage does not speciate in an interval is then $q=1-p$, and we assume that the probability that two speciation events occur in the same interval can be ignored (as it is of order $\mathcal{O}(\Delta t^2)$). The probability that there are $n$ lineages in the clade at the present day then is given by the probability that \emph{exactly} $n'=n-1$ speciation events occurred. Each time a speciation event occurs, the number of lineages --- and hence the total probability that a speciation event happens in the clade --- increases. As such, we must consider the probability that the $n'$ speciation events occur in the time intervals $j_1,j_2, \ldots, j_{n'}$, as well as the probability that speciation events do not occur between each of these times. For example, for the second speciation event to occur in interval $j_2$, neither of the lineages present can speciate between interval $j_{1}+1$ through interval $j_{2}-1$ --- an outcome which occurs with probability $q^{2(j_2-j_1-1)}$.
In intervals where speciation events occur (for example, $j_k$) any one of the $k$ species present can speciate, and all of the other $k-1$ species must not speciate. Therefore, the probability of observing a single speciation event in an interval is $k(pq^{k-1})$.  Carrying this forward, the resulting probability of observing $n$ species after $m$ intervals $\Pr(n|m)$ is:
\begin{equation}
\begin{aligned}
&\Pr(n|m)=\\
&\sum_{j_1=1}^{m-n'+1}\sum_{j_2=2}^{m-n'}\dots \sum_{j_{n'}=n'}^{m} \underbrace{q^{j_1-1}p}_{\text{\tiny{1\textsuperscript{st} speciation event}}}\underbrace{q^{2(j_2-j_1-1)}(2pq)}_{\text{\tiny{2\textsuperscript{nd} speciation event}}}\dots \underbrace{q^{n'(j_{n'}-j_{n'-1}-1)}(n'pq^{n'-1})}_{\text{\tiny{$n'$ speciation event}}}\underbrace{q^{n(m-j_{n'}-1)}}_{\text{\tiny{prob. of no more events}}}.
\end{aligned}
\end{equation}
\noindent where the sums are over all the possible intervals in which each subsequent speciation event can occur.

Given that we have assumed that $\Delta t=\frac{T}{m}$ is small, we are interested in the limit as $m\to\infty$. To evaluate this limit, we note that $p=\frac{\lambda T}{m}$ and $q=1-\frac{\lambda T}{m}$. Using the fact that $\displaystyle\lim_{m\to\infty}\left(1-\frac{\lambda T}{m}\right)^m=e^{-\lambda T}$, we can obtain the expression for the probabilities of clades of size 1 through 4 given in Table \ref{tab:SingleTypeProbabilities}. 

\begin{table}[H]
    \centering
    \renewcommand{\arraystretch}{1.25}
    \begin{tabular}{L{1cm} L{6cm} L{5cm}}
        \rowcolor{lightgray}
        $n$ & $\Pr(n\vert m)$ & $P_n(T)=\lim_{m\to\infty}\Pr(n\vert m)$\\
        \hline
        1 & $q^m$ & $e^{-\lambda T}$\\
        2 & $\frac{p q^{m-1} \left(q^m-1\right)}{q-1}$ & $e^{- \lambda  T} \left(1-e^{-\lambda  T}\right)$\\
        3 & $\frac{2 p^2 q^{m-2} \left(q^m-1\right) \left(q^m-q\right)}{(q-1)^2 (q+1)}$ & $e^{- \lambda  T} \left(1-e^{-\lambda  T}\right)^2$ \\
        4 & $\frac{6 p^3 q^{m-3} \left(q^m-1\right) \left(q^{2 m}-q^{m+1}-q^{m+2}+q^3\right)}{(q-1)^3 (q+1) \left(q^2+q+1\right)}$ & $e^{- \lambda  T} \left(1-e^{-\lambda  T}\right)^3$\\
    \end{tabular}
    \vspace{0.5cm}
    \caption{Probabilities of observing $n$ taxa after $m$ intervals and after time $T$ as the number of intervals goes to infinity. }
    \label{tab:SingleTypeProbabilities}
\end{table}
\noindent Calculating these expressions and limits for larger $n$ is arduous but they provided Yule (\citep[][p. 36]{Yule1924}) and us with an intuition as to the general expression for the probability of observing $n$ descendants of a single primordial lineage at time present day, $T$.
\begin{equation}
P_n(T)=e^{-\lambda T}\left(1-e^{-\lambda T}\right)^{n-1}.
\label{eq:PrimordialCladeSize}
\end{equation}
Using this probability, Yule (\citep[][p. 36]{Yule1924}) also calculated the expected number of descendants at time $T$:

\begin{equation}
\bar{n}(T)=E[n(T)]=\sum_{n=1}^\infty n e^{-\lambda T}\left(1-e^{-\lambda T}\right)^{n-1}=e^{\lambda T}.
\end{equation}
This important result showed that the average clade size grows as we would expect under a deterministic model.  Yule, surprisingly, did not consider the variance in $n(T)$ among independent realizations of the evolutionary process, which is given by:

\begin{equation}
\text{Var}(n(T))=E[(n(T))^2]-E[(n(T))]^2=e^{\lambda T}\left(e^{\lambda T}-1\right).
\end{equation}

\subsection*{Deriving Yule's result using stochastic processes}

In 1964, Bailey \citep{Bailey1964} re-derived Yule's core result above by conceptualizing diversification as a continuous-time stochastic process. As we discuss below, it was Bailey's formulation of the problem that influenced David Raup's pathbreaking work in paleobiology \citep{Raup1985}. We can express the Yule model in terms of a continuous time stochastic process with states $n=\{0,1,2,\dots\}$ representing the number of species present in the clade at time $t$.  The transition rate between state $i\in \mathbb{N}$ and state $j\in \mathbb{N}$ is then:

$$
q_{i,j}=\begin{cases}-\lambda i &j=i\quad \text{\textit{no speciation}}\\
\lambda i &j=i+1\quad \text{\textit{speciation}}\\
0 & \text{Otherwise}\end{cases}
$$

\noindent 
Using these rates we can then examine clade size using a master equation approach --- a system of ordinary differential equations describing the probability that a clade has $n$ species at time $t$:
\begin{equation}
\frac{dP_n(t)}{dt}=\sum_{j=0}^{\infty} q_{j,n} P_j(t)=-\lambda nP_n(t)+\lambda (n-1)P_{n-1}(t),
\end{equation}

\noindent where $P_{0}(t)=0$ (the probability the clade goes extinct) given that we assume there is at least one "primordial lineage" and no extinction (i.e., we only need to consider $n,i,\& j>0$ in the Yule model). Assuming, as Yule did, that there is a single primordial species the initial condition for this system gives:

$$
P_n(0)=\begin{cases}
1 & n=1\\
0 & \text{otherwise}
\end{cases}
$$

To solve this system of ODEs in two different ways.  We can either follow Yule and solve successively for the cases of small $n$, identify a pattern in the solutions and propose a general expression, or we can follow Bailey and obtain the general solution from generating functions (not shown here). While we think that Bailey's master equation approach is significantly easier to understand than Yule's probabilistic approach, it is hard to generalize to more complex diversification models as the system can be difficult or impossible to solve in these cases. The lesson evolutionary biology would learn in the 1980s from the development of Coalescent Theory \citep{Kingman1982} is that when faced with a hard problem, it is often easier to spin it around and solve it backwards.

\subsection*{Deriving Yule's result using a coalescent-style approach}

In our view, Nee and colleagues' \citep{NeeEtAl1992, NeeEtAl1994} formulation of the problem makes it even more straightforward to understand what Yule was doing than either Yule's own or Bailey's subsequent one. To see why, let $\tau$ be the time before the present day, such that $\tau=0$ is the present and $\tau=T$ is the time at which the clade originates.  Let $\mathcal{P}_n(\tau)$ be the probability that a single species present at time $\tau$ in the past gives rise to $n$ descendants at the present day.  In the manner of Nee \emph{et al.} \citep{NeeEtAl1994}, we can consider the possible events that can occur to a focal lineage going backward in time between $\tau$ and $\tau+\Delta \tau$.  Like Yule, we assume the time interval $\Delta\tau$ is small enough that the probability that two events occur in an interval can be neglected.  In this case there are only two outcomes, either nothing happens (with probability $1-\lambda \Delta\tau$), or a speciation event occurs (with probability $\lambda \Delta\tau$). Looking backward in time, a speciation event involves the joining of two lineages.  If we wish the ancestor of this speciation event to have $n$ offspring, then we must observe a speciation event that joins one lineage with $j\leq n$ descendants and another lineage with $n-j$ descendants \citep{LoucaEtAl2021}. As such,

\begin{equation}
\mathcal{P}_n(\tau+\Delta\tau)=\underbrace{\left(1-\lambda\Delta\tau\right)\mathcal{P}_n(\tau)}_{\text{nothing happens}}+\underbrace{\sum_{j=0}^n\lambda\Delta\tau\mathcal{P}_j(\tau)\mathcal{P}_{n-j}(\tau)}_{\text{speciation event}}+\mathcal{O}(\Delta\tau^2).
\end{equation}

\noindent Subtracting $\mathcal{P}_n(\tau)$ from both sides, dividing by $\Delta\tau$, and taking the limit as $\Delta\tau\to0$, we can use the definition of the derivative to obtain the system of backward-in-time master equations:

\begin{equation}
\frac{d}{d\tau}\mathcal{P}_n(\tau)=\lambda\mathcal{P}_n(\tau)+\sum_{j=0}^n\lambda\mathcal{P}_j(\tau)\mathcal{P}_{n-j}(\tau)
\end{equation}
These master equations have the convenient initial condition:
$$
\mathcal{P}_n(0)=\begin{cases}
1 & n=1\\
0 & \text{otherwise}
\end{cases}
$$
as each lineage observed at the present has, by definition, exactly one descendant. Noting that $\mathcal{P}_n(T)=P_n(T)$ (where $P_n(T)$ is defined as in the previous section), this system of ODEs can be solved recursively to obtain the distribution in Equation \eqref{eq:PrimordialCladeSize}.

\subsection*{Yule's empirical analyses and their interpretation}

Yule used the mathematical model to do a number of things. Among them was simply to compare the distribution of genera sizes generated by his model to those of empirical compilations of genera size for four taxonomic groups: snakes, lizards, and two groups of beetles. Yule was hesitant for the reader to over-interpret his results (“If any conclusions stated in this Introduction or in the body of the paper seem to be too confident, or at all dogmatic, I hope the reader will attribute the appearance to inadvertence of wording, or a simple desire to avoid the wearisome reiteration of qualifying phrases” [p. 32]). But he also cannot help but be impressed with how well his model matches empirical data:
\begin{quote}
    \emph{So far as the tests go I think it must be admitted that the formula given is capable of representing the facts with considerable precision, more closely indeed than we have any right to expect. One might well have expected the personal factor in classification, the practically cataclysmic destruction of species at numerous epochs in geological time, and all the varied changes that have diversified the organic history of our planet, to have left so many irregularities in the distribution that any formula could at most have given a very rough analogy with the general run of the data. But apparently the formula arrived at can do far more than this. [p. 58]}
\end{quote}
One interpretation of this result is that much of the variation in group sizes we observe can be attributed to stochasticity alone: one need not invoke the creator's inordinate fondness for any particular group of organisms. In thinking in these terms, Yule was more than fifty years ahead of his time (see for example, refs \citep{Schopf1979, SlowinskiGuyer1989}, discussed below) in appreciating the role of chance in generating major patterns of biodiversity. However, for reasons that we discuss in the section "The hierarchical nature of diversification", we should take this empirical result with a shaker of salt --- and the same applies to his attempt to address Willis' hypothesis regarding the correlation between age and geographic area that inspired this work in the first place. 

\section*{Yule's work had little impact on the early development of macroevolutionary biology}

Reading Yule’s paper today is a jarring experience. While the presentation of the work is certainly consistent with that of his time, the style of analysis and broader way in which he thinks of the problem seems right at home in the macroevolutionary literature of the late twentieth century. And this is consistent with the remarkable fate of Yule's paper; although it is widely recognized today as being a foundational contribution to the study of biodiversity, it had remarkably little influence on the development of our field for much of its history. 
For one, George Gaylord Simpson, the principal representative of paleontology in the modern synthesis and the pioneer of the study of taxonomic rates, did not seem even aware of Yule’s work. Not only is Yule not cited in either of his two influential books on macroevolution \citep{Simpson1984, Simpson1953}, Simpson does not place much emphasis on the importance of stochasticity or in developing null models for the sizes of the clades he is studying. 

In the 1970s, the study of paleontology underwent an incredible transformation --- this new "paleobiology" was quantitative and theory-driven. (We refer interested readers to David Sepkoski's riveting historical work \citep{Sepkoski2012} on the topic.) One of the key intellectual projects of the "paleobiological revolution" was that of the MBL group (named for their meetings at the Marine Biological Laboratory at Woods Hole). The MBL group --- a collaboration between paleobiologists David Raup, Stephen Jay Gould, Thomas Schopf, and the ecologist Dan Simberloff --- developed a series of simulation models, which were used to generate null distributions for the patterns of diversity under purely stochastic processes \citep{RaupEtAl1973, RaupGould1974, SchopfEtAl1975, GouldEtAl1977a}. They saw this work as a foundation for a "nomothetical" (law-forming) paleobiology; by comparing the patterns of idealized clades to those of the fossil record, they hoped to uncover a set of rules that could explain the ebb and flow of biodiversity \citep{Sepkoski2012}. Whether one judges their project a success or not, it undoubtedly did have lasting impact on paleobiology, as well as that of their intellectual descendants who primarily used molecular phylogenies, in orienting the fields around the idea that we should expect a wide range of outcomes from the processes of diversification and one needs to do statistical tests to compare observed data with these expectations. In retrospect, this work seems in line with how Yule was approaching similar problems in the 1920s, with the distinction that Yule was, like molecular phylogeneticists many years later, focused on a snapshot in time, rather than the complete history. Throughout the 1970s and early 1980s, the MBL researchers were unaware of Yule's work in this area (and indeed, we suspect that the MBL researchers could have saved some precious computing resources if they had known about Yule's analytical approaches for generating clade size distributions). As if to drive this omission home, Schopf wrote a lengthy paper \citep{Schopf1979} examining the history of thought on and study of stochasticity in generating macroevolutionary patterns. Schopf seemingly touches on every relevant work (as well as many only tangentially relevant ones) --- except that of Yule. And in a similar vein, Leigh van Valen \citep{vanValen1973} famously used the exponential distribution of species durations to argue for the central importance of Red Queen dynamics between hosts and parasites. However, exponential branch lengths also are expected to arise from the models of chance developed by Yule, yet van Valen does not consider this (see also \citep{Raup1975} on this point).

And this is not some quirk of Yule's paper being missed by a few key palentologists. David Kendall, who made a foundational contribution to the theory of birth-death processes \citep{Kendall1948}, and whose paper is now routinely cited in conjunction with Yule, seems unaware of his work (at least at the time he was first working on this problem, see ref. \citep{lambert2024}). (We believe that the original derivation of the birth-death process was by Feller \citep{feller1939grundlagen}, who also did not cite Yule.) Neither does Kendall mention Yule's contribution in his historical survey "Branching processes since 1863" \citep{Kendall1966}.  The mammalogist Sydney Anderson revisited the claims \citep{Anderson1974a} of Age and Area and developed his own stochastic model of clade evolution; despite providing a lengthy discussion of many previous attempts to explain the distribution of clade sizes, he only mentions Yule’s paper off-handedly and does not engage with the actual model or results. 

We do not have a good explanation for Yule’s pioneering work being overlooked for so long --- it was occasionally cited by prominent biodiversity scientists, including Preston \citep{Preston1948} and the above mentioned Anderson \citep{Anderson1974a} so it was not as if it was completely missed in these fields. And as noted in the introduction, Yule was well-known during his time and subsequently. We do think is it notable, however, that by the time macroevolution had come around to thinking about the problem of species diversity in a quantitative way that was reminiscent of Yule's approach, the theory of stochastic processes and the use of Monte Carlo simulations were well developed. It is only obvious in retrospect that Yule's model could be welded to these two to gain insight into the processes that have driven patterns of biodiversity. 

\section*{The rediscovery of Yule}

One macroevolutionary researcher who did eventually see the brilliance of Yule’s contribution was Raup \citep{German+2009+416+422}. As noted above, Raup was one of the key members of the MBL group, though he eventually diverged from the other members \citep{Sepkoski2012}. In his 1978 paper "Cohort analysis of generic survivorship" notes that "[a]n important but generally overlooked paleontological application [of branching processes] is that of Yule (1924)" (\citep{Raup1978}, p. 5); to the best of our knowledge, this is the first time Raup discussed Yule's work. Two of his contributions the prior year \citep{Raup1977, GouldEtAl1977a} on closely related topics did not mention Yule or Bailey's \citep{Bailey1964} text, which we suspect was where he first encountered Yule's work on the topic, given how extensively Raup leans on Bailey's work.
By the time he wrote his 1985 paper "Mathematical models of cladogenesis" \citep{Raup1985}, one of the most influential papers in the history of macroevolution, he prominently placed Yule's work as a foundational contribution in a long and illustrious intellectual lineage applying branching processes to biology. Following this work, researchers developed increasingly sophisticated rate estimators for fossil occurrence data \citep{Foote2000, Foote2003, SilvestroEtAl2014} and, as described above, Nee and colleagues \citep{NeeEtAl1994} demonstrated that diversification rates could be estimated from emerging dated molecular phylogenies. It is hard to overstate the importance of this contribution, not only in shaping modern diversification analyses (e.g., \citep{MagallonSanderson2001, MaddisonEtAl2007,AlfaroEtAl2009}) in macroevolution but also in epidemiology \citep{StadlerEtAl2012,MacPhersonEtAl2021a} and fields like cell biology \citep{StadlerEtAl2021}. 

It is important to emphasize however that this is only part of a broader and more complicated story. In the contemporary intellectual landscape, statistical phylogenetics and macroevolution are deeply intertwined fields \citep{o2012evolutionary, pennell2013integrative} . But this was not the case for most of the latter part of the twentieth century, where there were multiple distinct research traditions along similar themes that were not in close conversation with one another. So while we focus on Yule’s belated influence on macroevolutionary thought (i.e., on the research tradition that developed from paleobiology and extends to this day), his model also influenced the development of the first statistical models for reconstructing phylogenetic trees (see for instance, seminal contributions by  Edwards and Cavalli-Sfroza \citep{cavalli1967phylogenetic}, Edwards \citep{edwards1970estimation}, and Thompson \citep{thompson1975human} in the 1960s and early 1970s). And some mathematicians studying stochastic processes had recognized Yule’s model even earlier \citep{simon1955class, Bailey1964}. As these formerly disparate fields began to coalesce in the 1980s through the early twentieth century, there was a patchwork of historical traditions; in some of these Yule’s work was preeminent and in others, it was not. For example, in a 1989 paper \citep{SlowinskiGuyer1989}, the phylogenetic biologists Slowinksi and Guyer weave together many different threads on the history of using taxonomic and phylogenetic data to test for the effect of stochasticity in macroevolution, yet they leave Yule out of the picture entirely. In his influential 2001 paper \citep{aldous2001stochastic} on phylogenetic models for tree shape, Aldous discusses Yule at length but does not mention the contributions of Nee and colleagues \citep{NeeEtAl1992, NeeEtAl1994} more than half a decade earlier, nor does a recent summary by Fischer \citep{FischerEtAl2023a} of an extensive literature on tree balance statistics \citep[e.g.,][]{kirxpatrick1993searching, MooersHeard1997, BlumFrancois2006, AldousEtAl2011,Jones2011,  ColijnGardy2014, Henao-DiazPennell2023,FischerEtAl2023a,KerstingEtAl2024}. 

\section*{Contemporary questions in macroevolution Yule anticipated}

To close this Perspective, we wanted to highlight some additional issues and challenges that Yule raised in his 1925 paper that anticipated current debates in the field of macroevolution. As we show below, we do not think that his perspectives were always right but it is truly remarkable that he was engaged with these topics at a time when the field essentially did not yet exist.

\subsection*{The hierarchical nature of diversification}

\begin{quote}
\emph{The possible effect of size of genus (number of species in the genus) on the chance of a generic mutation is also ignored. This assumption may or may not be correct, but was deliberate. The generic characters are regarded as representing a main position of stability, and the chance of occurrence of a transfer from one main position of stability to another is regarded as independent of the number of minor positions of stability (species) which may have been taken up within the main position (genus) [p. 24].}
\end{quote}

We devoted our explanation of Yule's model to describing how he computed the probability distribution for sizes of clades of age $T$. This model, which forms the basis of much future work, was nested in a larger, justly forgotten model, even though we think the core idea was interesting. Recall that Yule was operating in the intellectual context where macromutationism was widely seen as a viable hypothesis. He deduced that there were two types of mutations: "generic mutations", large effect mutations that produced the synapomorphies that defined clades, and "specific mutations", which give rise to new speciation events. As he noted in the quote above, we considered these to be independent nested processes, perhaps akin to the processes that generated Simpson's adaptive zones \citep{Simpson1953} and the subsequent diversification within these. Today, most researchers studying lineage evolution consider all splits in the tree as representing speciation events, rather than as a combination of different types of events, including those that lead to higher taxonomic units. Paleobiologists often use the more inclusive term "origination rate" to describe the process of lineage splitting, as it is more agnostic regarding the taxonomic status of the newly formed groups \citep{Foote2000}, but nonetheless they only consider a single process. However, there have been some recent attempts to follow Yule's approach; i.e., to treat higher taxonomic groups as real evolutionary units and describe the processes by which they are formed \citep{Foote2012, MaruvkaEtAl2013, HumphreysBarraclough2014, HumphreysEtAl2016}.

But in the broader sense, Yule's conceptualization of diversification is a particular example of a broad category of models in which diversification occurs at different paces among different types  \citep{MacPhersonEtAl2021a}; in Yule's model, the two types are genera and species, which each have a distinct origination rate. The "multi-type models" most widely used today are "state-dependent speciation and extinction" (SSE) models pioneered by Maddison and colleagues \cite{MaddisonEtAl2007} which capture variation in diversification rates among lineages \citep{morlon2011reconciling, GoldbergIgic2012, BeaulieuOMeara2016, RasmussenStadler2019}. We can use the terminology of modern multi-type models to understand how Yule conceptualized diversification, how it stands in contrast to SSE models, and why many of his mathematical results are incorrect. We begin by noting three key differences between Yule's formulation of the problem and modern SSE models. First, Yule's state space (i.e., the number of types) was infinite --- while each species was labeled as belonging to a genus, there was no limit to the number of genera that could be present. Second, wile SSE models assign different evolutionary rates to each type in Yule's model, all species regardless of genera diversified at the rate $\lambda$. Third, since new genera were formed by a irreversible transition (\emph{sensu} \citep{GoldbergIgic2008a}) occurring at rate $\gamma$ in an existing lineage, this meant that most of the genera (except those that had no descendant genera) were paraphyletic. The widespread acceptance of the phylogenetic systematics approach to classification in the latter part of the twentieth century (see ref. \citep{hull2019science}) likely precluded the exploration of similar types of models.

For the single-type case, Yule calculated the probability that a genus has $n$ species given that it arose from a single "primordial" lineage $T$ time units ago.  The hierarchical multi-type model described above allowed Yule to consider a complementary measure of genus size, the probability that a \textit{randomly sampled genus} at time $T$ has $n$ species, which we denote as $\Pr(n|T)$. The sampled genus may be the primordial lineage itself or a newly formed genus arising from a generic mutation sometime between $t=0$ and $t=T$.  Averaging the genus size over the range of possible genus ages, $a$, we have by the law of total probability:
\begin{equation}
\Pr(n|T)=\int_0^T \Pr(a|T)\Pr(n|a,T)da,
\end{equation}
\noindent where $\Pr(n|a,T)$ is calculated as
\begin{equation}\label{eq:PrimordialGenusAge}
\Pr(n|a,T)=\Pr(n|a)=e^{-\lambda a}\left(1-e^{-\lambda a}\right)^{n-1}.
\end{equation}
Note that this assumes that $\Pr(n|a,T)$ is independent of the total age of the tree $T$. Equation \ref{eq:PrimordialGenusAge} is analogous to the probability calculated in Equation \ref{eq:PrimordialCladeSize}.

This leaves us to derive an expression for the probability that a randomly sampled genus at the present day $T$ arose at time $T-a$ and hence is of age $a$. Note that the age of a genus, $a$, cannot be older than the whole tree, $T$, hence the limits of the integral. If $T$ is small, then the sampled genus is likely to be the primordial genus, whereas if $T$ is large, most genera will have arisen since the origin of the tree, and $\Pr(a|T)$ will depend solely on when generic mutations occurred.  Recognizing this, Yule broke his analyses down into a transient case for small $T$ and a limiting case for large $T$.  We focus only on the limiting case here as the probability $\Pr(a|T)$ is in fact far more difficult to calculate than Yule proposed.

To understand how Yule proposed to calculate this quantity, let $m$ be the number of genera at the present day and $\tilde{m}$ the number of genera present at time $T-a$ when the focal genus arose.  Noting that the probability of a generic mutation occurring between time $T-a$ and $T-a+dt$ is $\gamma \tilde{m}\,dt$, Yule (\citep[][p.g 37]{Yule1924}) approximated this probability of a genus having age $a$ as:

$$
\Pr(a|T)=\frac{E[\gamma \tilde{m}]}{E[m]}
$$
\noindent where the expectations are over all stochastic realizations of the diversification process. There are several logical problems with this expression. First $m$ and $\tilde{m}$ and not independent; for example, trees with $\tilde{m}$ lineages at time $T-a$ in the past must have at least $m>\tilde{m}$ lineages at the present day. More subtly, the number of lineages at the present day is also not independent of whether or not a generic mutation occurred at time $T-a$. A modern backward-in-time approach could be used to calculate $\Pr(a|T)$ correctly, but doing so is neither useful, given the biological limitations of Yule's model, nor straightforward. We hope the reader will attribute our omission to a "simple desire to avoid the wearisome" derivations.    

\subsection*{Taxonomic patterns reflect both both biological realities and scientific conventions}

\begin{quote}
\emph{The desire of the systematist to break up a genus which he regarded as unwieldy might well tend to cause a deficiency of very large genera, but such a deficiency can hardly be held to be proved by the present tables [p. 58]}
\end{quote}

Here Yule is prescient about the potential confounding effect of what he elsewhere in the essay (quoted above) refers to as the "personal factor in classification” --- the fact that species counts per genus or the number of genera in a higher group are a consequence of both biology and human decisions --- even if he thought it was not a significant issue for his own analysis. Macroevolutionary biologists, both working with extinct \citep{Alroy2002,HendricksEtAl2014} and extant taxa \citep{Raikow1986, MinelliEtAl1991} have long shared Yule’s concern that inferences about diversification might be misled by biases in taxonomic practice. One might think that such issues become moot when looking at phylogenetic trees of lineages rather than Linnaean hierarchies --- indeed, this is one of the motivations for tree-based, rather than rank-based taxonomic systems \citep{deQueirozGauthier1994} --- but this is not the case \citep{FaurbyEtAl2016}. We necessarily look at subsets of the Tree of Life when conducting comparative analyses, and these subtrees often tend to be especially large or especially diverse ecologically, which might lead to biased inferences of typical rates \citep{PennellEtAl2012, LoucaEtAl2022a}; looking only at small trees has its downsides too \citep{BeaulieuOMeara2018}. And in practice, researchers tend to use taxonomic labels \citep{MagallonSanderson2001} to subset the tree. The nodes that we label with higher taxonomic classifications are viewed as special by systematicists (i.e., species that descend from a named node may harbor evolutionarily significant synapomorphies, for instance); the branch length distributions corresponding to trees stemming from named nodes may be different from random nodes on the Tree of Life \citep{Rabosky2009a, StadlerEtAl2014b, Henao-DiazPennell2023}. And furthermore, the problem of how to define terminal taxa never went away. This is particularly pernicious if the way we define lineages systematically biases the distribution of node ages. For example, bacterial “species” are typically defined using thresholds of sequence divergence and this can lead to a concentration of splitting events at certain times (see \citep{MorlonEtAl2012, LoucaEtAl2018}, for attempts to grapple with this bias). Another challenging situation is when taxa are defined inconsistently across an ecological gradient that we think might drive diversification. For example, a number of recent studies have found that recent speciation rates are faster in the temperate regions than the tropics \citep{SchluterPennell2017}, which goes against the predictions of classic theories for the existence of the latitudinal diversity gradient \citep{MittelbachEtAl2007a}. However, these findings are potentially undermined if biological species are often more narrowly defined in the better studied temperate regions \citep{FreemanPennell2021}. We anticipate that in the near future, researchers will be reckoning with how best to make inferences about the processes of diversification that are not so confounded with taxonomic practice.

\subsection*{The relationship between population dynamics and macroevolutionary rates}

\begin{quote}
\emph{At first, as the species spreads, the number of individuals must tend to increase. But over the very long periods which have to be considered there must be a countervailing tendency to ultimate decrease in the number of individuals, owing to the increase in the number of species. The area available being limited, the tendency, as it seems to me, must be towards greater and greater numbers of species and fewer individuals in each. Possibly the fact that ignoration of the number of individuals leads to results in close accordance with the data may only indicate this tendency first to increase and then to decrease in the size of the species [p. 24].}
\end{quote}

One notable aspect of reading Yule’s paper in light of contemporary research is how much space he devoted to justifying assuming a Poisson model to describe the speciation process; today, this assumption is overwhelmingly taken for granted and rarely justified, even though the macromutational model of speciation is no longer taken seriously as it was in Yule’s time. One way Yule justified this was pure ignorance; he figured that, in reality, speciation rates should be a function of population size, but did not know what type of function to specify. Yule’s intuition as to what this might look like is remarkably close to the canonical neutral theory of biodiversity and biogeography developed decades later \citep{Hubbell2001,Bell2001}. Neutral theory assumes that the number of individuals in an ecosystem is fixed, such that as the number of species in a clade increases, the average number of individuals per species must decrease, a concept known as the "zero-sum assumption" \citep{Hubbell2001,RosindellEtAl2011}. However, if one works out the implications of using such an assumption on the resulting phylogenetic trees, they look very different from those generated by Yule’s original model as well as empirical ones \citep{Nee2005}; some researchers argue that incorporating “protracted speciation”, where speciation happens as a prolonged process and not as an instantaneous point mutation, mitigates these failures \citep{RosindellEtAl2010}. We refer readers to Kopp \citep{Kopp2010} for a more in-depth discussion of the point mutation model of speciation in the context of neutral theory. We editorialize, though, that neutral theory has likely received more criticism about its speciation process than related approaches that tend not to make their assumptions explicit (e.g., standard birth-death models in macroevolution can be extended to include protracted speciation \citep{EtienneRosindell2012a, etienne2014estimating} but this is rarely considered in practice).

The neutral theory stands out as a rare attempt to actually model the relationship between speciation rate and population size. This is a particularly challenging problem since the rate limiting step for completing speciation is thought to be the persistence of nascent species \citep{RosenblumEtAl2012, DynesiusJansson2014, SchluterPennell2017, HarveyEtAl2019}, rather than the rate at which incipient species form; if this is the case, then it is not just the absolute population size that matters for speciation rates but also the population dynamics of these newly separated populations. Given this complexity, most of the current frontier of modeling approaches rely on simulations (e.g., refs. \citep{gascuel2015ecology, RosindellEtAl2015, ManceauEtAl2015, AguileeEtAl2018, RangelEtAl2018a}), which can allow speciation to emerge from complex eco-evolutionary dynamics, but when one wants to actually fit the model to data to test hypotheses about the causes of evolutionary rate variation, we must still typically resort to the standard branching processes that Yule was working with (but for exceptions, see \citep{PontarpEtAl2019, OvercastEtAl2021}). This is an obvious point for future development but given how little progress we have made in the last 100 years, one that will obviously be difficult.

\section*{Concluding remarks}

Yule's model is now firmly in the pantheon of core evolutionary models \citep{cohen2004mathematics}, sitting alongside the Wright-Fisher process \citep{Fisher1923,Wright1931}, the Moran process \citep{Moran1958}, and Kingman's coalescent \citep{Kingman1982}. As such, it hardly needs our promotion. Nonetheless, in this paper we have tried to make two cases. First, that it was far from inevitable that Yule would be viewed as a forebear of modern macroevolution; the field developed largely without his influence and his ingenuity was only belatedly recognized in this area (as we noted, it was recognized earlier in adjacent fields of study). Second, that Yule's insights into the \textit{process} of diversification were extraordinarily prescient given what was known about speciation biology in his time: the problems he was wrestling with as he was thinking through his model are still some of the most pressing challenges in modern macroevolution. We also tried to highlight some the ways that his conceptualization of the problem is distinct from our current ways of thinking and some limitations in his calculations. We hope that our paper is both a helpful guide and a little nudge for researchers who, like ourselves until embarrassingly recently, have left Yule's paper perpetually sitting on their "to read" pile.

\section*{Acknowledgements}

We thank David Jablonski, Michael Foote, Tom Archibald, and Ted Porter for providing historical insights. Noah Rosenberg, David Sepkoski, Joe Felsenstein, an anonymous review, Doc Edge, Melissa Guzman, and the members of PuRGE at the University of Idaho and the University of Wyoming provided helpful comments on the manuscript. This work was  supported by funds from the National Science and Engineering Research Council of Canada to A.M. (CRC-2021-00276 and RGPIN-2022-03113). 

\bibliography{support-files/bibliography_new}
\bibliographystyle{support-files/naturemag_natbib.bst}

\end{document}